\documentclass[a4paper,aps,prd,10pt,preprintnumbers,showpacs,twocolumn,superscriptaddress,nofootinbib,amsmath,amssymb]{revtex4-1}
\usepackage{graphicx}
\usepackage{cmap}
\usepackage[utf8]{inputenc}
\usepackage[T1]{fontenc}

\def\imo{i}

\def\K{{\cal K}}

\def\nq{\hspace*{-1em}}

\def\lal{&&\nq{}}

\def\beq{\begin{equation}}
\def\eeq{\end{equation}}
\def\bear{\begin{eqnarray}}
\def\bearr{\begin{eqnarray} \lal}
\def\ear{\end{eqnarray}}
\def\earn{\nonumber \end{eqnarray}}

\def\diag{\mathop{\rm diag}\nolimits}

\def\mn{_{\mu\nu}}
\def\MN{^{\mu\nu}}
\def\mN{_\mu^\nu}

\def\cF{{\mathcal F}}
\def\cL{{\mathcal L}}

\def\emag{electromagnetic}

\begin{document}
\title{Long Lived Quasinormal Modes of Regular and Extreme Black Holes}
\author{Milena Skvortsova}\email{milenas577@mail.ru}
\affiliation{Peoples' Friendship University of Russia (RUDN University), 6 Miklukho-Maklaya Street, Moscow, 117198, Russia}
\begin{abstract}
Recently, black hole models in a nonlinear modification of the Maxwell electrodynamics were suggested, possessing simultaneously properties of an extreme charge and regularity \cite{Bronnikov:2024izh}. We study quasinormal modes of a massive scalar field around such black holes and show that they are characterized by a comparatively small damping rate, indicating the possible existence of arbitrarily long-lived quasinormal modes, called {\it quasi-resonances}.     
\end{abstract}
\maketitle


\textbf{Introduction.} Quasinormal modes (QNMs) \cite{Kokkotas:1999bd,Berti:2009kk,Konoplya:2011qq} are fundamental oscillations that arise when black holes are perturbed, characterized by complex frequencies with real and imaginary components representing the oscillation frequency and the damping rate, respectively. These modes are crucial for understanding the stability and dynamic behavior of black holes, offering a unique window into the properties of spacetime and the nature of gravity. QNMs are pivotal in testing the predictions of General Relativity and have become a key observational tool with the advent of gravitational wave detectors like LIGO and Virgo \cite{LIGOScientific:2016aoc,LIGOScientific:2020zkf}. The study of QNMs extends across various black hole solutions, including those in modified theories of gravity and black holes with distinct charge and spin properties, helping to differentiate between theoretical models and uncovering the fundamental aspects of black hole physics.

Significant attention has been devoted to the quasinormal spectra of extreme and near-extreme black holes, where parameters such as electric charge or angular momentum reach their maximum values. The spectrum of the extremally charged Reissner-Nordström black hole is particularly noteworthy, as various perturbations can be related through supersymmetric transformations that maintain the extremal black hole's characteristics \cite{Onozawa:1995vu,Onozawa:1996ba}. Additionally, extremality introduces potential instability to the precisely extreme state, a phenomenon extensively discussed in the literature \cite{Aretakis:2013dpa,Aretakis:2011ha,Durkee:2010ea,Aretakis:2011gz,Lucietti:2012sf}. Furthermore, instability can also arise in highly charged near-extreme black holes \cite{Konoplya:2013sba}.

An intriguing aspect of certain black hole models is the potential to avoid the singularity at the origin, highlighting the limitations of General Relativity. The quasinormal modes of regular black holes have garnered considerable attention in recent years \cite{Konoplya:2024hfg,Bolokhov:2023ruj,Pedrotti:2024znu,Li:2016oif,Li:2022kch,Zhang:2024nny,Panotopoulos:2019qjk,Wahlang:2017zvk,Jha:2023wzo,Fernando:2012yw,Meng:2022oxg,Guo:2024jhg,Dubinsky:2024aeu,Flachi:2012nv}. Studies have shown that regular black holes can display unique features in their spectra, such as purely imaginary quasinormal modes or high sensitivity of overtones to the near horizon deformations \cite{Konoplya:2022hll}.

In \cite{Skvortsova:2024wly} we have considered quasinormal modes of massless scalar and Dirac fields for two black hole models that incorporate both extremality and regularity (with Minkowski and de Sitter cores), recently obtained in \cite{Bronnikov:2024izh}. Extreme charged black holes in the Einstein-Maxwell theory are singular; so that the regularity was achieved in \cite{Bronnikov:2024izh} through non-linear modifications of Maxwell electrodynamics. Here we will study the evolution of perturbations of a massive scalar field of the above two models, which qualitatively differs from the massless spectra both at the ringdown stage and at asymptotically late times.

Despite the short-range nature and observational challenges of massive fields, there are numerous reasons to study their perturbations. Consequently, the quasinormal modes of massive fields with various spins have been extensively explored \cite{Zhidenko:2006rs,Konoplya:2005hr,
Ohashi:2004wr,Zhang:2018jgj,Aragon:2020teq,Ponglertsakul:2020ufm,Gonzalez:2022upu,Ponglertsakul:2020ufm,Burikham:2017gdm,Sakalli:2022xrb}. These studies are motivated by several intriguing aspects and notable features that emerge in the spectrum when the mass term is varied.

First of all, the effective mass appears in perturbation equations of certain higher-dimensional gravity scenarios due to the action of the bulk on the brane \cite{Seahra:2004fg,Ishihara:2008re}. The massive graviton, whether in the context of massive gravity or as an effective term, can contribute to very long gravitational waves \cite{Konoplya:2023fmh}, which are currently being observed using the Pulsar Timing Array experiments \cite{NANOGrav:2023hvm}. Additionally, when a massless field is perturbed in the vicinity of a black hole surrounded by a magnetic field, an effective mass term can arise \cite{Konoplya:2007yy,Konoplya:2008hj,Wu:2015fwa,Davlataliev:2024mjl}. An interesting aspect of massive fields is their potential to produce arbitrarily long-lived frequencies at specific values of the field's mass \cite{Ohashi:2004wr,Konoplya:2004wg}. This phenomenon is quite broad, encompassing various field spins \cite{Fernandes:2021qvr,Konoplya:2017tvu,Percival:2020skc}, black hole backgrounds \cite{Konoplya:2006br,Zhidenko:2006rs,Zinhailo:2018ska,Bolokhov:2023bwm}, and even other types of compact objects such as wormholes \cite{Churilova:2019qph}. However, quasi-resonances do not exist in some cases \cite{Zinhailo:2024jzt,Konoplya:2005hr}, so their presence must be verified on a case-by-case basis.  Furthermore, the presence of a massive term significantly alters the late-time decay of the signal. Instead of the usual power-law tails that follow the quasinormal ringing, oscillatory tails emerge. These oscillatory tails have been thoroughly investigated in the literature \cite{Jing:2004zb,Koyama:2001qw,Konoplya:2006gq,Moderski:2001tk,Rogatko:2007zz,Koyama:2001ee,Dubinsky:2024jqi,Gibbons:2008gg,Dubinsky:2024hmn}.

In this paper, we study the quasinormal modes of a massive scalar field in the background of simultaneously extreme and regular black holes, as described in \cite{Bronnikov:2024izh}, within the framework of non-linear electrodynamics. To achieve this, we employ the higher-order WKB method and time-domain integration. The latter also enables us to analyze the asymptotic tails of these black holes, which differ from those of the Reissner–Nordström black holes.


\textbf{Extreme regular black holes and wave-like equation.} The structure of the Ricci and Einstein tensors can be represented by spherically symmetric  non-linear electrodynamics equations of motion.  For the Lagrangian $-\cL(\cF)$ we have
  $\cF = F\mn F\MN$, and $F\mn$ is the \emag\ field tensor, the SET is in general
\bearr                    \label{SET-F}
			T\mN = - 2 \mathcal{L_F} F_{\mu\sigma} F^{\nu\sigma}
					+(1/2) \delta\mN \mathcal{L(F)},
\ear
  with $\mathcal{L_F} = d\cL/d\cF$, and the \emag\ field equations are
		$	\nabla_\mu(\mathcal{L_F}F\MN) = 0.$
  Within spherically symmetric symmetry symmetry, we may consider only radial electric and magnetic
  fields. Then, following  \cite{Bronnikov:2024izh}, we have
\bearr         \label{T-F}		
		T\mN = (1/2) \diag(\cL,\ \cL,\ \cL - 4q^2 r^{-4} \cL_\cF,\
				\cL - 4q^2 r^{-4} \cL_\cF).\nonumber
\ear

For the case of extreme charged regular black holes with a de Sitter center, we have  
\begin{align}\nonumber
\mathcal{L}(\mathcal{F}) = & \ \frac{2 M (M r^4 + a^2 r^2 (-5 M + 6  \sqrt{a^2 + r^2}))}{(a^2 + r^2)^4} \nonumber \\
& + \frac{6 a^4 \sqrt{a^2 + r^2}}{(a^2 + r^2)^4},
\end{align}
while with a Minkowski center, we have  
\begin{align}\nonumber
\mathcal{L}(\mathcal{F}) = & \ \frac{2 M r (M r^5 + a^2 r^3 (-7 M + 8 r) + 16 a^4 r^2 + 8 a^6)}{(a^2 + r^2)^5}.\nonumber
\end{align}
Then, the Bronnikov extremally charged regular black holes in non-linear electrodynamics are described by the following metric  \cite{Bronnikov:2024izh}:
\begin{equation}\label{metric}
  ds^2=-f(r)dt^2+ f^{-1}(r) dr^2 +r^2(d\theta^2+\sin^2\theta d\phi^2),
\end{equation}
where the metric function \( f(r) \) differs based on the core type. For a regular black hole with a Minkowski core, the function is
$f_{1}(r)= \left(1-M r^3 (a ^2+r^2)^{-2}\right)^2,$
and for a black hole with a de Sitter core, it takes the form
$f_{2}(r)= \left(1-M r^2(a ^2+r^2)^{-3/2}\right)^2.$
In these expressions, \( a \) is a parameter signifying the degree of non-linearity, and \( M \) denotes the ADM mass. For convenience, we set \( M=1 \) to measure all dimensional quantities in units of mass. The existence of an event horizon is constrained by 
$a \lessapprox 0.32,$ for the Minkowski core model and by $a \lessapprox 0.38$  for the de Sitter core model.

\begin{figure*}
\resizebox{0.75 \linewidth}{!}{\includegraphics{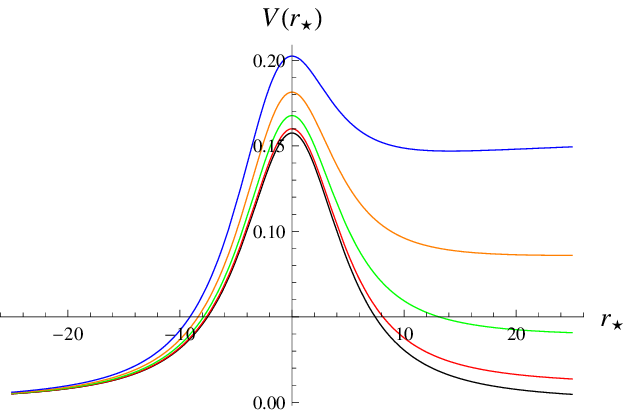}\includegraphics{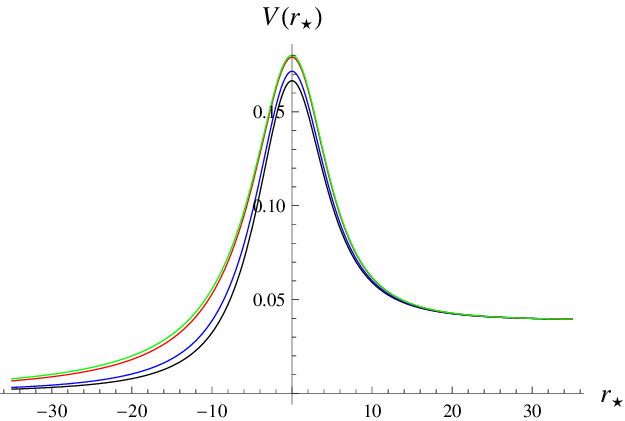}}
\caption{Left: Potential as a function of the tortoise coordinate of the $\ell=1$ scalar field for the Bronnikov extreme black hole with Minkowski core ($M=1$, $a=0.1$): $\mu=0$ (black), $\mu=0.1$ (red), $\mu=0.2$ (green), $\mu=0.3$ (orange), $\mu=0.4$ (blue). Right: Potentials for various $a$ ($a=0$ (black), $a=0.2$ (blue), $a=0.3$ (red), $a=0.31$ (orange)), $\mu=0.2$.}\label{fig:potentials}
\end{figure*}

The equations governing a massive scalar ($\Phi$) field in a general covariant framework are expressed as:
\begin{equation}\label{coveqs}
\partial_\mu \left(\sqrt{-g}g^{\mu \nu}\partial_\nu\Phi\right)= \mu^2 \sqrt{-g} \Phi. 
\end{equation}
We can use the following ansatz
\begin{equation}
\Phi(t, r, \theta, \phi) = e^{-i \omega t} Y_{\ell}(\theta, \phi) \frac{\Psi(r)}{r},
\end{equation}
Then, upon the separation of variables within the background metrics (\ref{metric}), these equations transform into a Schrödinger-like wave equation:
\begin{equation}\label{wave-equation}
(d^2/dr_*^2) \Psi+(\omega^2-V(r))\Psi=0,
\end{equation}
where the "tortoise coordinate" $r_*$ is defined as:
$dr_*\equiv dr/f(r),$
and the effective potential has the form
\begin{equation}\label{potentialScalar}
V(r)=f(r) \ell(\ell+1) r^{-2}+r^{-1} (d^2 r/dr_*^2) + f(r) \mu^2,
\end{equation}
where $\ell=0, 1, 2, \ldots$ represent the multipole numbers. The effective potentials for various values of the parameters $a$ and $\mu$ are depicted in fig. \ref{fig:potentials}. Notice, that $\mu$ in geometric units, together with the requirement $M=1$, means that $\mu \rightarrow \mu M/M_{Pl}^2$, where $M_{Pl}$ is the Planck mass.

\textbf{WKB method.} If the effective potential $V(r)$ in the wave equation (\ref{wave-equation}) takes the form of a barrier with a single peak, the WKB method proves effective in determining the dominant quasinormal modes, which must adhere to the boundary conditions:
$\Psi(r_*\to\pm\infty)\propto e^{\pm\imo \omega r_*},$
signifying purely ingoing waves at the horizon ($r \to-\infty$) and purely outgoing waves at spatial infinity ($r \to\infty$).

The WKB method entails matching asymptotic solutions, which satisfy the quasinormal boundary conditions, with a Taylor expansion around the peak of the potential barrier. The first-order WKB formula corresponds to the eikonal approximation and holds true in the limit $\ell \to \infty$. The general WKB expression for the frequencies is expanded around the eikonal limit as follows \cite{Konoplya:2019hlu}:
\begin{eqnarray}\label{WKBformula-spherical}
\omega^2&=&V_0+A_2(\K^2)+A_4(\K^2)+A_6(\K^2)+\ldots\\\nonumber
&-&\imo \K\sqrt{-2V_2}\left(1+A_3(\K^2)+A_5(\K^2)+A_7(\K^2)\ldots\right),
\end{eqnarray}
with the quasinormal mode matching conditions given by
$\K=n+ (1/2), \quad n=0,1,2,\ldots,$
where $n$ denotes the overtone number, $V_0$ is the maximum value of the effective potential, $V_2$ is the second derivative of the potential at this maximum with respect to the tortoise coordinate, and $A_i$ (for $i=2, 3, 4, \ldots$) are the WKB correction terms beyond the eikonal approximation, dependent on $\K$ and higher-order derivatives of the potential at its maximum up to order $2i$. The explicit forms of $A_i$ can be found in \cite{Iyer:1986np} for the second and third WKB orders, in \cite{Konoplya:2003ii} for the fourth to sixth orders, and in \cite{Matyjasek:2017psv} for the seventh to thirteenth orders. At orders higher than six the general form of the corrections is too cumbersome and cannot be explicitly written out.  This WKB approach for determining quasinormal modes has been extensively applied at various orders in numerous studies (see, for instance, \cite{Konoplya:2006ar,Kokkotas:2010zd,Balart:2023odm,DuttaRoy:2022ytr,Kodama:2009bf,Al-Badawi:2023lvx,Gonzalez:2022ote,Chen:2023akf,Panotopoulos:2017hns,Rincon:2018sgd,Panotopoulos:2018pvu}), including recent analyses by the author \cite{Skvortsova:2023zmj,Skvortsova:2024atk} of quasinormal modes of $2+1$ dimensional \cite{Konoplya:2020ibi} and quantum corrected \cite{Lewandowski:2022zce} black holes. Here,  as well as in the above previous works, we mostly use the 6th order WKB method with $\tilde{m}=4$
Pade approximants as prescribed in \cite{Matyjasek:2017psv,Konoplya:2019hlu}, where $\tilde{m}$ determines the structure of the Pade approximants and is defined in \cite{Matyjasek:2017psv}. However, we see that at large values of the coupling parameter $a$, the 13th WKB order with   $\tilde{m}=7$ is sometimes slightly more accurate. 


\textbf{Time-domain integration.} The accuracy of the aforementioned WKB method can be confirmed by comparing it with an independent approach using time-domain integration. For time-domain integration, we utilized the Gundlach-Price-Pullin discretization scheme \cite{Gundlach:1993tp}:
\begin{eqnarray}
\Psi\left(N\right)&=&\Psi\left(W\right)+\Psi\left(E\right)-\Psi\left(S\right)\nonumber\\
&&- \Delta^2V\left(S\right)\frac{\Psi\left(W\right)+\Psi\left(E\right)}{8}+{\cal O}\left(\Delta^4\right),\label{Discretization}
\end{eqnarray}
where the integration points are defined as follows: $N\equiv\left(u+\Delta,v+\Delta\right)$, $W\equiv\left(u+\Delta,v\right)$, $E\equiv\left(u,v+\Delta\right)$, and $S\equiv\left(u,v\right)$. This method has been extensively used in numerous studies \cite{Konoplya:2014lha,Konoplya:2020jgt,Qian:2022kaq,Varghese:2011ku,Momennia:2022tug,Melis:2024kfr} and has proven to be sufficiently accurate. To extract the values of frequencies from the time-domain profile, we employ the Prony method, which fits the profile data with a sum of damped exponents:
\begin{equation}\label{damping-exponents}
\Psi(t)\simeq\sum_{i=1}^pC_ie^{-i\omega_i t}.
\end{equation}
We assume that the quasinormal ringing stage begins at some initial time $t_0=0$ and ends at $t=Nh$, where $N\geq2p-1$. The relation (\ref{damping-exponents}) is then satisfied for each point of the profile:
\begin{equation}
x_n\equiv\Psi(nh)=\sum_{j=1}^pC_je^{-i\omega_j nh}=\sum_{j=1}^pC_jz_j^n.
\end{equation}
We determine $z_i$ from the known values of $x_n$ and then calculate the quasinormal frequencies $\omega_i$. Quasinormal modes are typically extracted from time-domain profiles when the ringdown stage includes a sufficient number of oscillations. The higher the multipole number $\ell$, the longer the ringdown period.

\begin{figure*}
\resizebox{0.90 \linewidth}{!}{\includegraphics{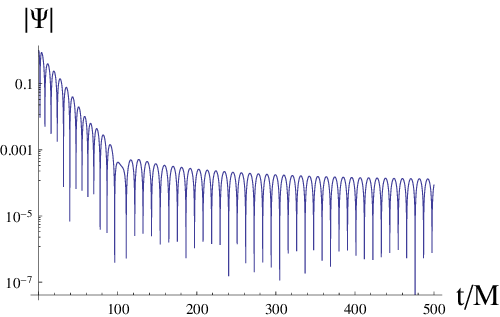}\includegraphics{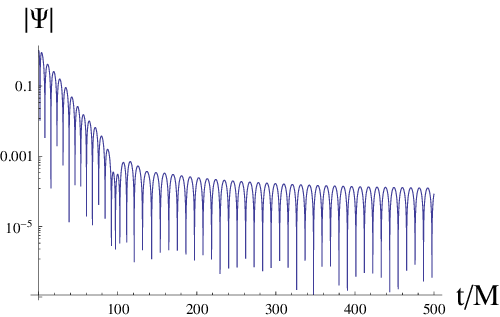}\includegraphics{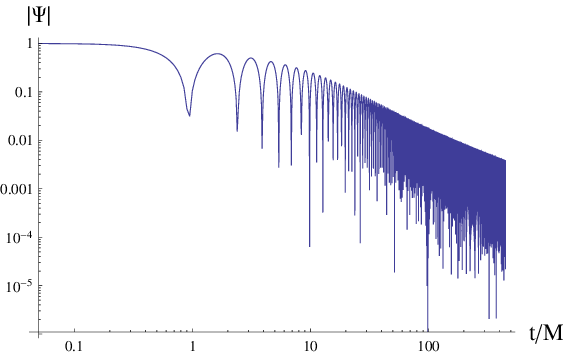}}
\caption{Time-domain profile in the-semi logarithmic plot for a massive scalar perturbations $\ell=1$, $\mu=0.3$  hole $a = 0$ (left) and $a = 0.3$ (middle). Logarithmic plot for  $\ell=1$, $\mu=3$ $a = 0.3$ (right).}\label{fig:timedomain}
\end{figure*}

\begin{table*}
\begin{tabular}{c c c c c c}
\hline
$\mu $ & Prony fit & WKB6 Padé & rel. error $\mathrm{Re}(\omega)$ & rel. error $\mathrm{Im}(\omega)$ & $\ell$ \\
\hline
$0$ & $0.133319-0.096030 i$ & $0.134025-0.095165 i$ & $0.530\%$ & $0.901\%$ & $0$ \\
$0.05$ & $0.135586-0.094180 i$ & $0.135096-0.093669 i$ & $0.361\%$ & $0.542\%$ & $0$ \\
$0.1$ & $0.139846-0.085470 i$ & $0.137668-0.089325 i$ & $1.56\%$ & $4.51\%$ & $0$ \\
$0.15$ & $0.160902-0.080667 i$ & $0.141272-0.082483 i$ & $12.2\%$ & $2.25\%$ & $0$ \\
$0.2$ & $0.187910-0.071119 i$ & $0.150933-0.072500 i$ & $19.7\%$ & $1.94\%$ & $0$ \\
$0.25$ & $0.230967-0.064792 i$ & $0.182909-0.056867 i$ & $20.8\%$ & $12.2\%$ & $0$ \\
$0.3$ & $0.260165-0.061461 i$ &  &  &  & $0$ \\
$0.35$ & $0.297382-0.061654 i$ &  &  &  & $0$ \\
$0.4$ & $0.332401-0.059580 i$ &  &  &  & $0$ \\
$0.45$ & $0.373326-0.061814 i$ &  &  &  & $0$ \\
$0.5$ & $0.409790-0.060949 i$ &  &  &  & $0$ \\
$0.55$ & $0.447103-0.064489 i$ &  &  &  & $0$ \\
\hline
$0$ & $0.377649-0.089333 i$ & $0.377642-0.089382 i$ & $0.00184\%$ & $0.0556\%$ & $1$ \\
$0.05$ & $0.378315-0.088995 i$ & $0.378390-0.089111 i$ & $0.0199\%$ & $0.130\%$ & $1$ \\
$0.1$ & $0.380532-0.087982 i$ & $0.380644-0.088292 i$ & $0.0294\%$ & $0.352\%$ & $1$ \\
$0.15$ & $0.384076-0.086630 i$ & $0.384447-0.086744 i$ & $0.0966\%$ & $0.131\%$ & $1$ \\
$0.2$ & $0.389316-0.084544 i$ & $0.389645-0.084735 i$ & $0.0847\%$ & $0.225\%$ & $1$ \\
$0.25$ & $0.395912-0.081689 i$ & $0.396461-0.081944 i$ & $0.139\%$ & $0.313\%$ & $1$ \\
$0.3$ & $0.404683-0.077225 i$ & $0.404842-0.078296 i$ & $0.0391\%$ & $1.39\%$ & $1$ \\
$0.35$ & $0.416027-0.070794 i$ & $0.414793-0.073638 i$ & $0.296\%$ & $4.02\%$ & $1$ \\
$0.4$ & $0.431704-0.065443 i$ & $0.426271-0.067777 i$ & $1.26\%$ & $3.57\%$ & $1$ \\
$0.45$ & $0.453919-0.060200 i$ & $0.439162-0.060471 i$ & $3.25\%$ & $0.451\%$ & $1$ \\
$0.5$ & $0.480033-0.057488 i$ & $0.453263-0.052163 i$ & $5.58\%$ & $9.26\%$ & $1$ \\
$0.55$ & $0.511693-0.053217 i$ & $0.471285-0.031677 i$ & $7.90\%$ & $40.5\%$ & $1$ \\
\hline
\end{tabular}
\caption{Comparison of the quasinormal frequencies for the regular black hole with a Minkowski core obtained by time-domain integration and the 6th order WKB approach with Padé approximants for $a=0$, and $\ell=0,1$ ($M=1$).}
\label{MassiveMerged}
\end{table*}

\begin{table*}
\begin{tabular}{c c c c c c}
\hline
$\mu $ & Prony fit & WKB6 Padé & rel. error $\mathrm{Re} (\omega)$ & rel. error $\mathrm{Im} (\omega)$ & $\ell$ \\
\hline
$0$ & $0.136270-0.091219 i$ & $0.136690-0.091031 i$ & $0.308\%$ & $0.206\%$ & $0$ \\
$0.05$ & $0.138568-0.089574 i$ & $0.138418-0.090138 i$ & $0.108\%$ & $0.630\%$ & $0$ \\
$0.1$ & $0.143405-0.081629 i$ & $0.141503-0.084935 i$ & $1.33\%$ & $4.05\%$ & $0$ \\
$0.15$ & $0.162625-0.077472 i$ & $0.145326-0.079024 i$ & $10.6\%$ & $2.00\%$ & $0$ \\
$0.2$ & $0.189395-0.068681 i$ & $0.153857-0.070010 i$ & $18.8\%$ & $1.94\%$ & $0$ \\
$0.25$ & $0.230955-0.062806 i$ & $0.184459-0.054463 i$ & $20.1\%$ & $13.3\%$ & $0$ \\
$0.3$ & $0.260308-0.060200 i$ &  &  &  & $0$ \\
$0.35$ & $0.297349-0.060221 i$ &  &  &  & $0$ \\
$0.4$ & $0.331994-0.058654 i$ &  &  &  & $0$ \\
$0.45$ & $0.372738-0.060926 i$ &  &  &  & $0$ \\
$0.5$ & $0.409163-0.060705 i$ &  &  &  & $0$ \\
$0.55$ & $0.446986-0.064521 i$ &  &  &  & $0$ \\
\hline
$0$ & $0.396451-0.081250 i$ & $0.396459-0.081303 i$ & $0.00210\%$ & $0.0652\%$ & $1$ \\
$0.05$ & $0.396995-0.080966 i$ & $0.397120-0.081114 i$ & $0.0315\%$ & $0.183\%$ & $1$ \\
$0.1$ & $0.398989-0.080324 i$ & $0.399101-0.080541 i$ & $0.0281\%$ & $0.271\%$ & $1$ \\
$0.15$ & $0.401709-0.079418 i$ & $0.402426-0.079550 i$ & $0.178\%$ & $0.167\%$ & $1$ \\
$0.2$ & $0.406478-0.077835 i$ & $0.407108-0.078095 i$ & $0.155\%$ & $0.334\%$ & $1$ \\
$0.25$ & $0.412442-0.075615 i$ & $0.413177-0.076099 i$ & $0.178\%$ & $0.640\%$ & $1$ \\
$0.3$ & $0.420233-0.072320 i$ & $0.420662-0.073454 i$ & $0.102\%$ & $1.57\%$ & $1$ \\
$0.35$ & $0.430736-0.067029 i$ & $0.429583-0.070006 i$ & $0.268\%$ & $4.44\%$ & $1$ \\
$0.4$ & $0.444480-0.062427 i$ & $0.439930-0.065551 i$ & $1.02\%$ & $5.00\%$ & $1$ \\
$0.45$ & $0.463489-0.058819 i$ & $0.451675-0.059872 i$ & $2.55\%$ & $1.79\%$ & $1$ \\
$0.5$ & $0.486446-0.056300 i$ & $0.464511-0.052845 i$ & $4.51\%$ & $6.14\%$ & $1$ \\
$0.55$ & $0.515607-0.053756 i$ & $0.482775-0.042640 i$ & $6.37\%$ & $20.7\%$ & $1$ \\
\hline
\end{tabular}
\caption{Comparison of the quasinormal frequencies for the regular black hole with de Sitter core obtained by time-domain integration and the 6th order WKB approach with Padé approximants for $a=0.3$, and $\ell=0,1$ ($M=1$).}
\label{checkMassiveMerged}
\end{table*}

\begin{figure*}
\resizebox{1.0 \linewidth}{!}{\includegraphics{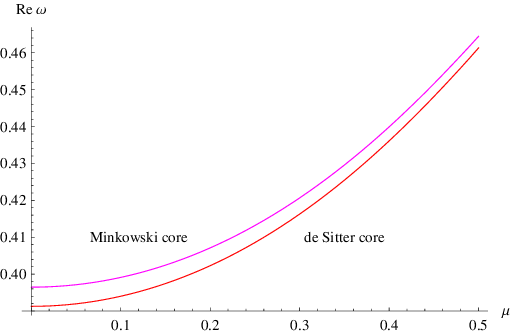}\includegraphics{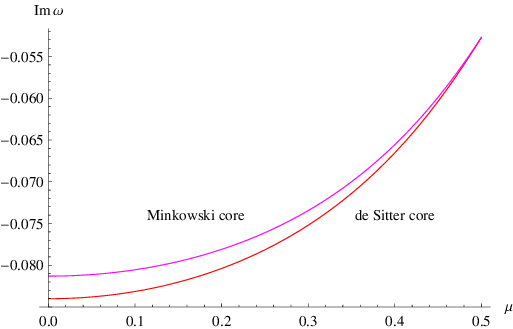}\includegraphics{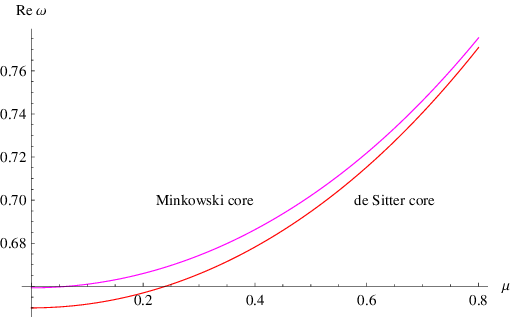}\includegraphics{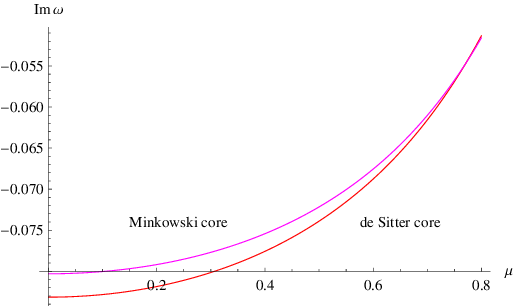}}
\caption{Quasinormal modes for the massive scalar perturbations: $\ell=1$ (first and second), and $\ell=2$ (thrid and forth) $a =0.3$. Purple color is for the Minkowski core and red is for de Sitter core.}\label{fig:QNM1plot}
\end{figure*}

\textbf{Quasinormal modes.} Quasinormal modes are calculated here with the higher order WKB method and time-domain integration. The first method can be applied to relatively small values of the field's mass $\mu$, because at larger $\mu$ the effective potential does not have peak. Nevertheless, for small $\mu \leq 0.3$ both methods agree very well as can be seen, for instance, in tables I-II for the regular black hole with Minkowski core. For larger $\mu$ we should trust to the time-domain integration.

However, this method cannot be applied to sufficiently large values of $\mu$ for extracting the frequencies. In contrast to the regime of small and moderate $\mu$, which exhibits a distinct stage of quasinormal ringing that can be clearly separated from the asymptotic tails (see Fig. \ref{fig:timedomain}), the case of large $\mu$ presents a different behavior. Specifically, for large $\mu$, the ringdown phase blends with the late-time tail, making it difficult to distinguish the two. From a technical perspective, the Prony fitting procedure yields significantly different frequencies for slightly varying fitting intervals, indicating a lack of convergence.

For the Schwarzschild case, the quasinormal spectrum of a massive field is known to exhibit peculiar behavior \cite{Ohashi:2004wr,Konoplya:2004wg}.  
The damping rate of the least damped, fundamental mode decreases, approaching zero at a certain critical value of $\mu$. Beyond this point, the fundamental mode disappears from the spectrum, and the first overtone takes its place. This process repeats as $\mu$ increases further, with the first overtone undergoing the same transition. As a result, the spectrum of massive fields can contain arbitrarily long-lived quasinormal modes, resembling standing waves. However, as mentioned earlier, this phenomenon is not universal, and there are configurations where it does not occur \cite{Zinhailo:2024jzt,Konoplya:2005hr}. 

In this paper, we demonstrate clear indications that a similar phenomenon occurs for extreme regular black holes within the framework of non-linear electrodynamics.

We can see that the parameter $a$ almost does not change 
the real oscillation frequency, but noticeably suppresses the damping rate.
This behaviour takes place for null and small $\mu$, but at moderate mass of the field the difference between frequencies corresponding to various values of $a$
disappear, meaning that relatively small effect of the non-linear electrodynamics is suppressed by the dominating massive term. From fig. \ref{fig:QNM1plot} it follows that the damping rate at larger $\mu$ does not depend on $a$ and goes to the Schwarzschild limit, for which arbitrarily long lived modes take place at some critical $\mu$, as it is known from precise calculations made by Leaver method \cite{Ohashi:2004wr,Konoplya:2004wg}. Thus, while we cannot achieve the true regime of quasi-resonances with WKB or time-domain integration methods, we achieve the regime where the damping rate is indistinguishable from the Schwarzschild case. The precise results on quasinormal modes could also be achieved via using other convergent methods, such as various varaints of the pseudospectral method \cite{Becar:2024agj,Stashko:2024wuq,Konoplya:2024lch} or Bernstein polynomial method \cite{Fortuna:2020obg,Konoplya:2022zav}.

At later times, following the quasinormal ringing period  we observe the intermediate types of tails, which depend on the couplings $a$ and multipole number $\ell$. Example of such tails can be seen in figs. \ref{fig:timedomain} for the black hole model with the Minkowski core.  
At  times 
$t/M > (\mu M)^{-3},$
we observe asymptotic tails corresponding to the regime $t \rightarrow \infty$,  which can clearly be seen in right plot on fig. \ref{fig:timedomain}.  The decay law is:
\begin{equation}
|\Psi| \sim t^{-1} \sin (A(\mu) t), \quad t \rightarrow \infty.
\end{equation}
Here $A(\mu)$ is some function, which, could be found via fitting numerical data for various values of $\mu$.  
The intermediate and asymptotic tails are oscillatory with power-law envelope. This decay law differs from that of  \cite{Koyama:2000hj,Koyama:2001qw} for the singular non-extreme RN black holes. The late-time tails are typically governed by the asymptotic behavior of the effective potential. Consequently, for other theories involving non-linear electrodynamics, extra fields, modifications of gravity or different ((anti)-de Sitter) asymptotics \cite{Al-Badawi:2024jnt,Toshmatov:2019gxg,Waeber:2024ilt,Ahmed:2024qeu,Sakalli:2016jkf,Al-Badawi:2024jnt}, this should be examined separately.

All of the aforementioned aspects of late-time decay were illustrated for the black hole model with a Minkowski core. A similar behavior is observed for black holes with a de Sitter core, whose quasinormal modes are shown in Figs. \ref{fig:QNM1plot}. It is evident that both models can be distinguished, as the black holes with a Minkowski core exhibit higher oscillation frequencies and damping rates.

For higher $\ell$ and larger $\mu$, we noticed that the higher-order WKB method shows slightly better agreement with time-domain integration. For instance, for $a=0$ and $\ell=1$, the results are as follows: (a) Prony fit $\omega = 0.511693 - 0.053217 i$, (b) 13th order WKB $\omega = 0.469383 - 0.043615 i$, and (c) 6th order WKB $\omega = 0.471285 - 0.031677 i$.


\textbf{Conclusions.} We have studied  quasinormal modes of a massive scalar field perturbations around regular, maximally charged black holes recently discovered within the framework of nonlinear modifications of Maxwell electrodynamics \cite{Bronnikov:2024izh}, allowing for Minkowski and de Sitter cores. Calculations were performed using two independent methods: the higher order WKB method and time-domain integration, yielding a good agreement between them in the common range of applicability. We have shown that there are long lived modes in the spectrum, indicating the existence of arbitrarily long-lived quasinormal modes - quasi-resonances. At late times the ringdown is changed by the intermediate and asymptotic oscillatory tails with power-law envelopes. 

An interesting question, beyond the scope of our study, is how different astrophysical environments influence the observed quasinormal modes of massive fields. For the simplest case of a Schwarzschild black hole, this problem was recently investigated in \cite{Konoplya:2024wds,Davlataliev:2024mjl}. However, to the best of our knowledge, no such studies exist for scenarios involving non-linear electrodynamics.

Another interesting direction for future study is the extension of this work to higher-dimensional black holes. The quasinormal mode spectrum in higher dimensions exhibits several distinctive features, such as purely imaginary modes and potential instabilities \cite{Cardoso:2003vt,Cardoso:2003qd,Konoplya:2017ymp}.

\textbf{Acknowledgments.} 
This work was supported by RUDN University research project FSSF-2023-0003.

\bibliographystyle{apsrev4-1}
\bibliography{bibliography}

\end{document}